\documentclass[11pt,a4paper]{article}
\usepackage{hyperref}
\usepackage[utf8]{inputenc}
\usepackage{latexsym,graphicx,amssymb,amsmath}
\usepackage{cite}
\usepackage{slashed}
\usepackage{bm}
\usepackage{float}
\usepackage{authblk}
\usepackage{geometry}
\usepackage{color}
\newcommand{\bea}{\begin{eqnarray}}
\newcommand{\eea}{\end{eqnarray}}
\newcommand{\be}{\begin{equation}}
\newcommand{\ee}{\end{equation}}
\begin{document}

\title{\vspace{-1cm}\textbf {On the masses of light pseudoscalar mesons}}
\author[a]{Chang-Yong Liu
\thanks{liuchangyong@nwsuaf.edu.cn}}
\author[b]{Wei He
\thanks{weihephys@foxmail.com}}
\affil[a]{College of Science, Northwest A\&F University, Yangling, Shaanxi
712100, China}
\affil[b]{School of Physics and Astronomy, China West Normal University,
Nanchong 637002, China}
\date{}
\maketitle
\begin{abstract}

We investigate the masses of light pseudoscalar mesons by the method based on a new anomaly free condition for axial vector current. By this viewpoint, the field theories discussed here do not have the $U(1)$ problem. We calculate the masses of nine light pseudoscalar mesons, with theoretical result agrees reasonably good with experiment.
\\
\begin{description}


\item[Keywords:] Pseudoscalar mesons, Pauli-Villars regularization, Axial anomaly

\end{description}
\end{abstract}




\section{Introduction}

\begin{figure}[H]
\center
\includegraphics[width=0.3\textwidth]{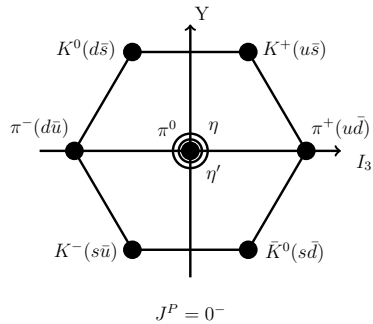}
\caption{\label{m} The nine pseudoscalar mesons. }
\end{figure}

In modern physics, hadrons are considered as bound states of quarks and gluons, there is no satisfying first-principle method to determine the mass spectrum of light hadron because the strong coupling nature of the problem.  At low energy effective theory,  the mesons are treated as Goldstone boson of the broken chiral $SU(3)_L \times SU(3)_R$ symmetry, in that way some  phenomenological relation of meson masses can be obtained \cite{Weinberg:1978kz, Gasser:1984gg, Pich:1995bw, Scherer:2002tk, Nambu:1961tp, Nambu:1961fr, Klevansky:1992qe, Meisinger:1995ih, Costa:2008dp}.
Concerning the nature of baryons in this picture, one approach is to view baryons as soliton solutions of the effective theory of meson fields.
In this paper, we study the mass of some light mesons by a method developed in the previous works by the first author \cite{Liu:2021pmt,Liu:2022zdo}. The particle content of the pseudoscalar mesons are shown in Figure.  \ref{m}.

For the light quarks $u, d, $ and $s$, there is the approximate $SU(3)_L \times SU(3)_R \times U(1)_B \times U(1)_A$ symmetry since the constitute quark mass parameters are small compared to the quantum chromodynamics energy scale  \cite{Weinberg:1996kr, Langacker:2017uah}. The  the quark sector of the Lagrangian density is
\bea
\mathcal{L}&=&\bar{q} \left(i D\!\!\!\!/-M^q \right)q \nonumber \\
&=&\bar{q}_L i D\!\!\!\!/ \ q_L+\bar{q}_R i D\!\!\!\!/ \ q_R-\left(\bar{q}_LM^q q_R+\bar{q}_RM^q q_L\right),
\eea
where $q=(u\  d\  s)$ is the quark fields, $M^q={\rm{diag}}(m_u\ m_d\ m_s)$ is the diagonal quark mass matrix, the color indices are suppressed.
The pseudoscalar meson is connected with the composite pseudoscalar field operator $\pi^i$ which is
\bea
\pi^i\equiv -i \bar{q} \frac{\lambda^i}{2} \gamma^5 q,
\eea
for $i=0\cdots 8$, where $\lambda^0=\sqrt{\frac{2}{3}}I$ ($I$ is the $3\times 3$ identity matrix), the other $\lambda^i$s are the Gell-Mann matrices. Then the $\pi^0$ is $SU(3)$ singlet and the others are each identified with an octets of $SU(3)$. The axial vector currents $A_{5\mu}^i$ is defined by
\bea
A_{5\mu}^i=\bar{q} \gamma_{\mu}\frac{\lambda^i}{2} \gamma^5 q.
\eea
It is connected with the composite pseudoscalar field operators $\pi^i$ through the partially conserved axial current (PCAC) relation \cite{Gell-Mann:1964hhf}, which  states that the divergences of the axial-vector currents are proportional to the renormalized field
 operators representing the lowest-lying pseudoscalar octet, that is
\bea
\label{Ai}
 \partial^{\mu} A_{5\mu}^i=f_\pi m^2_\pi\pi^i.
\eea
This relation relates the broken approximate symmetry to the mass of corresponding Nambu-Goldstone boson.
The symmetry is not exact because of the Adler-Bell-Jackiw (ABJ) anomaly \cite{Adler:1969gk,Bell:1969ts,Bilal:2008qx}.

\begin{figure}[h]
\center
\includegraphics[width=0.4\textwidth]{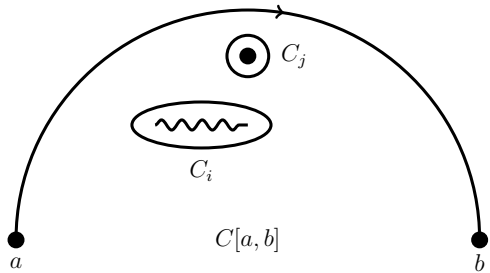}
\caption{\label{integral} A smooth contour $C[a,b]$ in complex plane staring from $a$ to $b$.  }
\end{figure}

Then, one can use ABJ anomaly to study the masses of pseudoscalar particles. In spite of success of the effective field theories, these theories always have the $U(1)$ problem \cite{Weinberg:1975ui,Witten:1979vv,Veneziano:1979ec}. In previous works by the first author, a new anomaly free condition for chiral anomaly is used to study the masses of some neutral pseudoscalar mesons \cite{Liu:2021pmt,Liu:2022zdo}.
In this paper, we will generalize this method to other pseudoscalar mesons.
The idea is as follows. In calculating diagrams that leads to anomalous terms using the Feynman parameter representation,
the integrand contains poles and branch cuts. Usually, the integral is understood as the Cauchy principal value of real integral.
However, one can treat the computation as complex variable integral and take into account the contribution of poles and branch cuts.
Suppose the function $f(z)$ has poles and branch cuts, as in Figure \ref{integral}, then the integral of $f(z)$ along the contour $C[a,b]$ can be expressed as \cite{Liu:2018dfm}
\bea
\label{fz}
\int_{C[a,b]}f(z)dz=P\int_a^bf(z)dz+\sum n_i \oint_{C_i}f(z)dz,
\eea
where the $C_i$ is a closed curve circling a pole or a branch cut. The term $P\int_a^bf(z)dz$ denotes the principal value which takes value in the main single-valued branch. The number $n_i\in \mathbb{Z}$ denotes contour encircles $n_i$ times around the pole or branch cut.
The second term in (\ref{fz}) gives an extra term that could be used to cancel the usual anomalous term.

\par

The paper is organized as follows. In Section 2, we use a new anomaly free condition to find the mass formula of neutral mesons $\pi^0$, $\eta$ and $\eta'$. In Section 3, we obtain the mass formula of charged light mesons $\pi^{\pm}$ and $K^{\pm}$. In Section 4, we find the mass formula of mesons $K^0$ and $\bar{K}^0$. In Section 5, we compare our results with the experimental values. We end with the conclusions.

\section{The mass formula of neutral mesons $\pi^0$, $\eta$ and $\eta'$}

\begin{figure}[H]
\center
\includegraphics[width=0.6\textwidth]{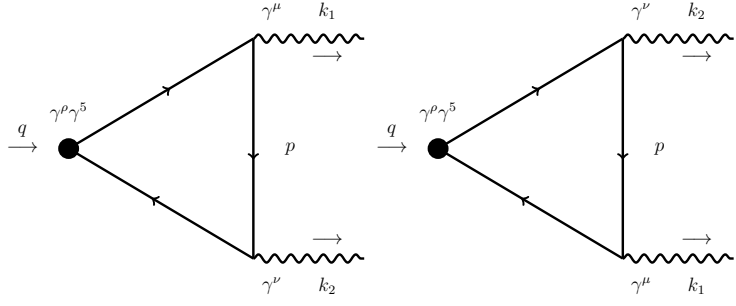}
\caption{\label{triangle} Triangle diagrams contributing to chiral anomaly.  }
\end{figure}

The neutral mesons $\pi^0$, $\eta$ and $\eta'$ are bound states of quarks and anti-quarks of the same flavor, or more precisely superpositions of quark-anti-quark pairs. We consider the electro-magnetic interaction in 3+1 dimensions with Lagrangian density
\bea
\mathcal{L}=-\frac{1}{4}F_{\mu\nu}F^{\mu\nu}+\bar{\psi}(i\partial\!\!\!/-e_fA\!\!\!/-m)\psi,
\eea
where the fermion can be the quark components of a meson, $e_f$ is the charge and $m$ is the mass. The vector and the axial vector currents are \cite{Peskin:1995ev,Itzykson:1980rh}
\bea  \label{vector}
j^{\mu}=\bar{\psi}\gamma^{\mu}\psi,\quad j^{\mu 5}=\bar{\psi}\gamma^{\mu}\gamma^{5}\psi.
\eea
The pseudoscalar mesons are related to the anomalous axial vector currents as in the relation (\ref{Ai}). To study the chiral anomaly, one needs to analyze the matrix element
\bea
\int d^4x {\rm{e}}^{-iqx}\langle k_1,k_2|j^{\rho 5}(x)|0\rangle=(2\pi)^4 \delta^{4}(k_1+k_2-q)\epsilon^{\ast}_{\mu}(k_1)\epsilon^{\ast}_{\nu}(k_2)M_5^{\rho\mu\nu}(k_1,k_2,m),
\eea
where the tensor $M_5^{\rho\mu\nu}(k_1,k_2,m)$ is computed by diagrams in Figure \ref{triangle},
\bea
M_5^{\rho\mu\nu}(k_1,k_2,m)=e_f^2\int\frac{d^4p}{(2\pi)^4}{\rm{Tr}}[\frac{i}{p\!\!\!/-m}\gamma^{\mu}\frac{i}{p\!\!\!/+k\!\!\!/_1-m}
\gamma^{\rho}\gamma^{5}\frac{i}{p\!\!\!/-k\!\!\!/_2-m}\gamma^{\nu}
+\left(
  \begin{array}{c}
    \mu\leftrightarrow\nu \\
    k_1\leftrightarrow k_2 \\
  \end{array}
\right)].
\eea
\par
\begin{figure}[H]
\center
\includegraphics[width=0.8\textwidth]{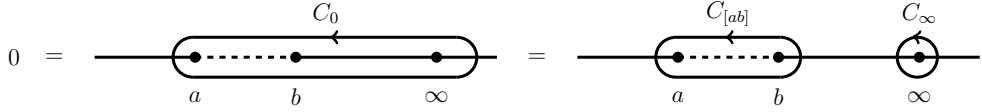}
\caption{\label{circle} The contour integral relation.  }
\end{figure}

In momentum space, taking the divergence of the axial current is equivalent to dotting with $iq^{\rho}$, then the anomaly of current is proportional to
\bea \label{aaa}
iq_{\rho}M_5^{\rho\mu\nu}(k_1,k_2,m)=e_f^2\int\frac{d^4p}{(2\pi)^4}{\rm{Tr}}[\frac{p\!\!\!/+m}{p^2-m^2}\gamma^{\mu}\frac{p\!\!\!/+k\!\!\!/_1+m}{(p+k_1)^2-m^2}
q\!\!\!/\gamma^{5}\frac{p\!\!\!/-k\!\!\!/_2+m}{(p-k_2)^2-m}\gamma^{\nu}
+\left(
  \begin{array}{c}
    \mu\leftrightarrow\nu \\
    k_1\leftrightarrow k_2 \\
  \end{array}
\right)].
\eea
The integral can be regulated by introducing a set of Pauli-Villars fields \cite{Pauli:1949zm} with masses $M_j$ and compute
\bea \label{TTD}
iq_{\rho}\widetilde{M}_5^{\rho\mu\nu}(k_1,k_2,m)=iq_{\rho}M_5^{\rho\mu\nu}(k_1,k_2,m)-i\sum_{j}c_j q_{\rho}M_5^{\rho\mu\nu}(k_1,k_2,M_j).
\eea
To remove the divergences, one can impose the conditions
\bea
\label{condition}
\sum_{j}c_j=1, \quad \sum_{j}c_j M_j^2=m^2.
\eea
Taking the limit $M_j\rightarrow \infty$, the subtracted tensor in (\ref{TTD}) becomes

\bea
iq_{\rho}\widetilde{M}_5^{\rho\mu\nu}(k_1,k_2,m)&=&\frac{m^2e_f^2}{2\pi^2}\epsilon^{\mu\nu}_{\quad \rho\sigma}k_1^{\rho}k_2^{\sigma}\int_0^1 dx \int_0^{1-x}dy \frac{1}{m^2-x k_1^2-y k_2^2+(x k_1-y k_2)^2}\nonumber \\
&-&\frac{e_f^2}{4\pi^2}\epsilon^{\mu\nu}_{\quad \rho\sigma}k_1^{\rho}k_2^{\sigma}
+\left(
  \begin{array}{c}
    \mu\leftrightarrow\nu \\
    k_1\leftrightarrow k_2 \\
  \end{array}
\right).
\eea
At this point, we apply the integration formula (\ref{fz}) to compute $iq_{\rho}\widetilde{M}_5^{\rho\mu\nu}(k_1,k_2,m)$.
We consider a general case that the external fields are $k_1^2\neq 0$, $k_2^2\neq 0$ and $q^2=(k_1+k_2)^2=2k_1\cdot k_2+k_1^2+k_2^2$.
Using the formula (\ref{fz}) to compute the integral of the $y$-variable, since the integrand only contains poles, we obtain
\bea
&&\int_0^{1-x}dy \frac{1}{m^2-x k_1^2-y k_2^2+(x k_1-y k_2)^2}=P\int_0^{1-x}dy \frac{1}{m^2-xk_1^2+x^2k_1^2-y k_2^2-xy(q^2-k_1^2-k_2^2)+y^2k^2_2}\nonumber \\
&&+\frac{2\pi n_1 i}{\sqrt{\left(x(q^2-k_1^2-k_2^2)+k^2_2\right)^2-4(m^2-xk_1^2+x^2k_1^2)k^2_2}},
\label{section2Yintegrated}
\eea
where $n_1$ is an integer, $n_1\in \mathbb{Z}$.  One can substitute $q^2-k_1^2-k_2^2=2k_1\cdot k_2$ in the formula, but here we keep it. There remains the integral of the $x$-variable. The second term in (\ref{section2Yintegrated}) as the integrand has a branch cut and a pole at infinity, applying the integration formula (\ref{fz}), besides the principle part there is a integral that can be recast in the form
\bea
\oint_{C_{[ab]}} \frac{dz}{\sqrt{(z-a)(z-b)}},
\eea
where $C_{[ab]}$ is the contour around the branch cut $[a,b]$ as shown in Figure \ref{circle}.
One can deform the integral contour from $C_{[ab]}$ to $C_{\infty}$, by the relation
\bea
0=\oint_{C_0} \frac{dz}{\sqrt{(z-a)(z-b)}}=\oint_{C_{[ab]}} \frac{dz}{\sqrt{(z-a)(z-b)}}+\oint_{C_{\infty}} \frac{dz}{\sqrt{(z-a)(z-b)}},
\eea
the integral encircling the branch cut is
\bea
\label{cab}
\oint_{C_{[ab]}} \frac{dz}{\sqrt{(z-a)(z-b)}}=-\oint_{C_{\infty}} \frac{dz}{\sqrt{(z-a)(z-b)}}=2\pi i.
\eea
\par
\noindent
Therefore, we obtain the integral of $x$-variable for the second term as
\bea
&&\int_0^1 dx \frac{1}{\sqrt{\left(x(q^2-k_1^2-k_2^2)+k^2_2\right)^2-4(m^2-xk_1^2+x^2k_1^2)k^2_2}}\nonumber \\
&=&P\int_0^1 dx \frac{1}{\sqrt{\left(x(q^2-k_1^2-k_2^2)+k^2_2\right)^2-4(m^2-xk_1^2+x^2k_1^2)k^2_2}}\nonumber \\
&-&\frac{2 \pi i n_2}
{\sqrt{(q^2-k_1^2-k_2^2)^2-4 k_1^2 k_2^2}},
\eea
where $n_2$ is another integer, $n_2\in \mathbb{Z}$. After these manipulations of integrals, the subtracted tensor can be expressed as
\bea
\label{mm5}
&&iq_{\rho}\widetilde{M}_5^{\rho\mu\nu}(k_1,k_2,m)=\frac{m^2e_f^2}{2\pi^2}\epsilon^{\mu\nu}_{\quad \rho\sigma}k_1^{\rho}k_2^{\sigma}\left[ P \int_0^1 dx \int_0^{1-x}dy \frac{1}{m^2-x k_1^2-y k_2^2+(x k_1-y k_2)^2}\right.\nonumber \\
&&\left. +P\int_0^1 dx \frac{2\pi n_1 i}{\sqrt{\left(x(q^2-k_1^2-k_2^2)+k^2_2\right)^2-4(m^2-xk_1^2+x^2k_1^2)k^2_2}}+ \right.\nonumber \\
&&\left.\frac{(2 \pi)^2 n_1 n_2}
{\sqrt{(q^2-k_1^2-k_2^2)^2-4 k_1^2 k_2^2}}\right]
-\frac{e_f^2}{4\pi^2}\epsilon^{\mu\nu}_{\quad \rho\sigma}k_1^{\rho}k_2^{\sigma}
+\left(
  \begin{array}{c}
    \mu\leftrightarrow\nu \\
    k_1\leftrightarrow k_2 \\
  \end{array}
\right).
\eea
The last two terms on the right side of (\ref{mm5}) are anomalous contributions, since they have opposite signs, there is a chance to cancel the anomaly by imposing the following relation on external momentums,
\bea
\label{mm6}
\sqrt{(q^2-k_1^2-k_2^2)^2-4 k_1^2 k_2^2}=8\pi^2 m^2 n_1n_2=8 n \pi^2 m^2, \quad n=n_1n_2 \in Z.
\eea
The on shell $q^2$ is related with the mass of meson $m_P$ as $q^2=m_P^2$. Substituting the relation (\ref{mm6}) into (\ref{mm5}), we obtain
\bea
\label{mm7}
&&iq_{\rho}\widetilde{M}_5^{\rho\mu\nu}(k_1,k_2,m)=\frac{m^2e_f^2}{2\pi^2}\epsilon^{\mu\nu}_{\quad \rho\sigma}k_1^{\rho}k_2^{\sigma}\left[ P \int_0^1 dx \int_0^{1-x}dy \frac{1}{m^2-x k_1^2-y k_2^2+(x k_1-y k_2)^2}\right.\nonumber \\
&&\left. +P\int_0^1 dx \frac{2\pi k i}{\sqrt{\left(x(q^2-k_1^2-k_2^2)+k^2_2\right)^2-4(m^2-xk_1^2+x^2k_1^2)k^2_2}}\right]+
\left(
  \begin{array}{c}
    \mu\leftrightarrow\nu \\
    k_1\leftrightarrow k_2 \\
  \end{array}
\right).
\eea
When taking the massless limit $m\rightarrow 0$ for the fermion, using $\frac{q^2}{m^2}\rightarrow \infty$ one obtains the limit for the expression (\ref{mm7}),
\bea
\lim_{m\rightarrow 0} iq_{\rho}\widetilde{M}_5^{\rho\mu\nu}(k_1,k_2,m)=0,
\eea
the axial vector current is free of anomaly, the anomalous term is cancelled because of the integration formula (\ref{fz}) and the imposed condition (\ref{mm6}).

Specific to the neutral mesons $\pi^0$, $\eta$ and $\eta'$, we need to use the axial vector currents which are connected with the these meson field
operators. The meson is a superposition state of light quark-anti-quark pairs, the meson field operator can be written as
\bea
 {\pi^0}^5=\alpha\bar{u}\gamma^{5}u+\beta\bar{d}\gamma^{5}d+\gamma\bar{s}\gamma^{5}s,
\eea
where $\alpha,\beta,\gamma$ are coefficients for meson wavefunction, their values would be assigned in section \ref{ComparingWithExperimentalValues} to compute the mason mass,
then the corresponding axial-vector current is
\bea
{j^0}^{\mu 5}=\alpha\bar{u}\gamma^{\mu}\gamma^{5}u+\beta\bar{d}\gamma^{\mu}\gamma^{5}d+\gamma\bar{s}\gamma^{\mu}\gamma^{5}s.
\eea
To accommodate the fact that the meson is composed of a few quark flavors, the charge $e_f^2$ in formula (\ref{mm5}) should be substituted by $\alpha e_u^2+\beta e_d^2+\gamma e_s^2$, likewise the quantity $m^2e_f^2$ should be substituted by $\alpha m_u^2 e_u^2+\beta m_d^2 e_d^2+\gamma m_s^2 e_s^2$, therefore the mass parameter in (\ref{mm6}) is substituted as in the following,
\bea
\label{qqu}
\sqrt{(q^2-k_1^2-k_2^2)^2-4 k_1^2 k_2^2}=8\pi^2n\frac{\alpha m_u^2 e_u^2+\beta m_d^2 e_d^2+\gamma m_s^2 e_s^2}{\alpha e_u^2+\beta e_d^2+\gamma e_s^2},
\eea
where the charge of quarks are $e_u=\frac{2}{3}e$, $e_d=-\frac{1}{3}e$ and $e_s=-\frac{1}{3}e$, the value of quark masses $m_u, m_d, m_s$ are given in section \ref{ComparingWithExperimentalValues}.
To calculate the meson masses using formula (\ref{qqu}), one needs also the value of $k_1$ and $k_2$, a  momentum takes zero value if the corresponding vector boson is on shell, otherwise it is determined by equation of anomaly free conditions, as we explain in sections \ref{MassOfChargedMesons} and \ref{MassesKandKBAR}.
For meson at the ground state the integer takes value $n=1$, for excited state it takes value of larger integers.
The calculation of various light meson masses is presented in section \ref{ComparingWithExperimentalValues}, comparing with our previous work \cite{Liu:2022zdo}, the formula (\ref{qqu}) gives more accurate meson masses.

\section{The mass formula of charged light mesons $\pi^{\pm}$ and $K^{\pm}$}\label{MassOfChargedMesons}

\begin{figure}[H]
\center
\includegraphics[width=0.8\textwidth]{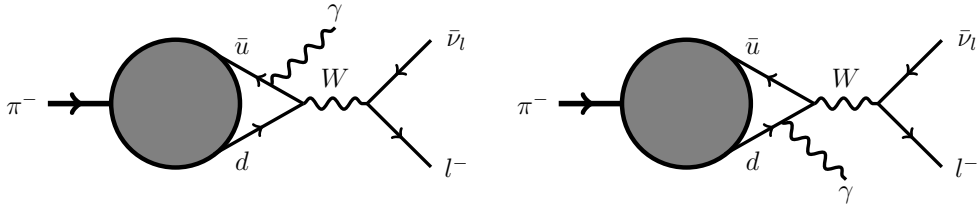}
\caption{\label{pi} The decay channels of $\pi^{-} \rightarrow l^{-}\bar{\nu}_{l}\gamma$.  }
\end{figure}

\begin{figure}[H]
\center
\includegraphics[width=0.6\textwidth]{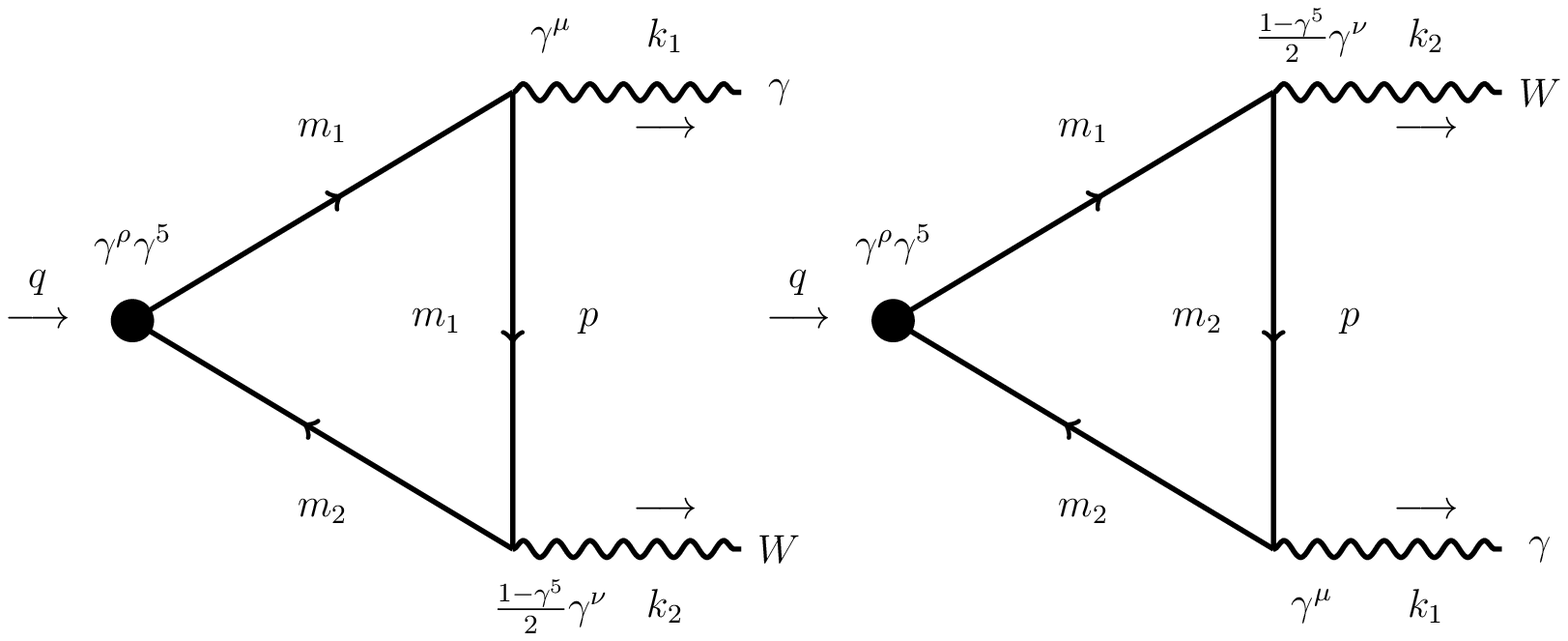}
\caption{\label{triangle2} Triangle diagrams contributing to calculate the divergence of current $j^{\rho5}=i\bar{\psi}_1\gamma^{\rho}\gamma^5 \psi_2$.  }
\end{figure}

The charged light pseudoscalar mesons like $\pi^{\pm}$ and $K^{\pm}$ are composed of quark-anti-quark of different flavors. The axial vector currents corresponding to these mesons have the form $j^{\rho5}=i\bar{\psi}_1\gamma^{\rho}\gamma^5 \psi_2$, where $\psi_1$ and $\psi_2$ represent quarks of different flavors. This axial vector currents suffer from the ABJ anomalies too. For example, the decay modes of the $\pi^{-}$ include the processes $\pi^{-} \rightarrow \l^{-}\bar{\nu}_{l}\gamma$, with $l$ a lepton, as in Figure \ref{pi} where the leptons are produced through an intermediate heavy gauge boson of weak interaction. In the diagrams there are subdiagrams which are equivalent to those shown in Figure \ref{triangle2}, they are related to the ABJ anomaly calculation. Notice that there is a chirality projection operator $(1-\gamma_5)/2$ associated with the $W$-boson. Since the $W$-bosons are off-shell, one shall take $k_2^2\neq 0$.
\par

\par

Following the same procedure as in the previous section, the triangle diagrams in (\ref{triangle2}) lead to the following tensor \cite{Weinberg:1996kr,Langacker:2017uah},
\bea
&&{M^{\pm}}_{5}^{\rho\mu\nu}(k_1,k_2,m_1,m_2)=N_c\frac{e_1g_WV_{12}}{2\sqrt{2}}\int\frac{d^4p}{(2\pi)^4}{\rm{Tr}}\left[\frac{i}{p\!\!\!/-m_1}\gamma^{\mu}\frac{i}{p\!\!\!/+k\!\!\!/_1-m_1}
i\gamma^{\rho}\gamma^{5}\frac{i}{p\!\!\!/-k\!\!\!/_2-m_2}(1-\gamma^5)\gamma^{\nu}\right] \nonumber \\
&&+N_c\frac{e_2g_WV_{12}}{2\sqrt{2}}\int\frac{d^4p}{(2\pi)^4}{\rm{Tr}}\left[\frac{i}{p\!\!\!/-m_2}(1-\gamma^5)\gamma^{\nu}\frac{i}{p\!\!\!/+k\!\!\!/_2-m_1}
i\gamma^{\rho}\gamma^{5}\frac{i}{p\!\!\!/-k\!\!\!/_1-m_2}\gamma^{\mu}\right],
\eea
where $N_c$ is the number of colors, $e_i$ ($i=1,2$) is the charge carried by quark field $\psi_i$, $m_i$ ($i=1,2$) is the mass of $\psi_i$, and $g_W$ is the weak coupling constant and $V_{12}$ is the element of the Cabibbo-Kobayashi-Maskawa (CKM) matrix $V_{{\rm{CKM}}}$ \cite{Cabibbo:1963yz,Kobayashi:1973fv,Peskin:1995ev,Weinberg:1996kr,Thomson:2013zua},
\bea
\begin{pmatrix}
V_{ud}& V_{us} & V_{ub}\\
V_{cd}& V_{cs} & V_{cb}\\
V_{td}& V_{ts} & V_{tb}
\end{pmatrix}
=\begin{pmatrix}
0.97370\pm 0.00014& 0.2245\pm 0.0008 & 0.00382\pm 0.00024\\
0.221\pm0.004& 0.987\pm0.011 & 0.0410\pm 0.0014\\
0.0080\pm 0.0003& 0.0388\pm 0.0011 & 1.013\pm 0.030
\end{pmatrix}.
\eea
Taking the massless limit $m_1\to 0$, $m_2$ $\rightarrow$ 0, the anomaly free condition leads to the following equations,
\bea
\begin{cases}
  iq_{\rho}{M^{\pm}}_{5}^{\rho\mu\nu}(k_1,k_2,m_1,m_2)\big{|}_{\epsilon}=0, \\
   ik_{2\nu}{M^{\pm}}_{5}^{\rho\mu\nu}(k_1,k_2,m_1,m_2)\big{|}_{\epsilon}=0,\\
    ik_{1\mu}{M^{\pm}}_{5}^{\rho\mu\nu}(k_1,k_2,m_1,m_2)\big{|}_{\epsilon}=0,
 \end{cases}
\eea
where the mark $\big{|}_{\epsilon}$ indicates the terms containing the Levi-Civita symbol $\epsilon^{\mu\nu\rho\sigma}$ factor.
The $ik_{1\mu}{M^{\pm}}_{5}^{\rho\mu\nu}(k_1,k_2,m_1,m_2)\big{|}_{\epsilon}=0$ is satisfied because of the $U(1)$ gauge symmetry, so we need to  consider only the quantities $iq_{\rho}{M^{\pm}}_{5}^{\rho\mu\nu}(k_1,k_2,m_1,m_2)\big{|}_{\epsilon}$ and $ik_{2\nu}{M^{\pm}}_{5}^{\rho\mu\nu}(k_1,k_2,m_1,m_2)\big{|}_{\epsilon}$.

The $\epsilon$-dependent piece comes from the anomaly term $i{M^{\pm}}_{5,a}^{\rho\mu\nu}(k_1,k_2,m_1,m_2)$,
\bea
&&{M^{\pm}}_{5,a}^{\rho\mu\nu}(k_1,k_2,m_1,m_2)=N_c\frac{e_1g_WV_{12}}{2\sqrt{2}}\int\frac{d^4p}{(2\pi)^4}{\rm{Tr}}\left[\frac{i}{p\!\!\!/-m_1}\gamma^{\mu}\frac{i}{p\!\!\!/+k\!\!\!/_1-m_1}
i\gamma^{\rho}\gamma^{5}\frac{i}{p\!\!\!/-k\!\!\!/_2-m_2}\gamma^{\nu}\right] \nonumber \\
&&+N_c\frac{e_2g_WV_{12}}{2\sqrt{2}}\int\frac{d^4p}{(2\pi)^4}{\rm{Tr}}\left[\frac{i}{p\!\!\!/-m_2}\gamma^{\nu}\frac{i}{p\!\!\!/+k\!\!\!/_2-m_1}
i\gamma^{\rho}\gamma^{5}\frac{i}{p\!\!\!/-k\!\!\!/_1-m_2}\gamma^{\mu}\right].
\eea
Dotting with $iq^{\rho}$ yields
\bea \label{aaa2}
&&iq_{\rho}{M^{\pm}}_{5,a}^{\rho\mu\nu}(k_1,k_2,m_1,m_2)=iN_c\frac{e_1g_WV_{12}}{2\sqrt{2}}
\int\frac{d^4p}{(2\pi)^4}{\rm{Tr}}\left[\frac{p\!\!\!/+m_1}{p^2-m_1^2}\gamma^{\mu}\frac{p\!\!\!/+k\!\!\!/_1+m_1}{(p+k_1)^2-m_1^2}
q\!\!\!/\gamma^{5}\frac{p\!\!\!/-k\!\!\!/_2+m_2}{(p-k_2)^2-m_2^2}\gamma^{\nu}\right] \nonumber \\
&&+iN_c\frac{e_2g_WV_{12}}{2\sqrt{2}}
\int\frac{d^4p}{(2\pi)^4}{\rm{Tr}}\left[\frac{p\!\!\!/+m_2}{p^2-m_2^2}\gamma^{\nu}\frac{p\!\!\!/+k\!\!\!/_2+m_1}{(p+k_2)^2-m_1^2}
q\!\!\!/\gamma^{5}\frac{p\!\!\!/-k\!\!\!/_1+m_2}{(p-k_1)^2-m_2^2}\gamma^{\mu}\right].
\eea
Then using the following relation,
\bea
q\!\!\!/\gamma^{5}=(k\!\!\!/_1+k\!\!\!/_2)\gamma^{5}=(p\!\!\!/+k\!\!\!/_1-m_1)\gamma^{5}+\gamma^{5}(p\!\!\!/-k\!\!\!/_2-m_2)+(m_1+m_2)\gamma^{5},
\eea
we obtain
\bea
&&iq_{\rho}{M^{\pm}}_{5,a}^{\rho\mu\nu}(k_1,k_2,m_1,m_2)=\nonumber \\
&&-N_c\frac{e_1g_WV_{12}}{2\sqrt{2}}4(m_1+m_2)\epsilon^{\mu\nu}_{\quad \rho\sigma}\int\frac{d^4p}{(2\pi)^4}\frac{(m_2-m_1)k_1^{\rho}p^{\sigma}+m_1k_1^{\rho}k_2^{\sigma}}{[p^2-m_1^2][(p+k_1)^2-m_1^2][(p-k_2)^2-m_2^2]}\nonumber \\
&&-N_c\frac{e_2g_WV_{12}}{2\sqrt{2}}4(m_1+m_2)\epsilon^{\mu\nu}_{\quad \rho\sigma}\int\frac{d^4p}{(2\pi)^4}\frac{(m_2-m_1)k_1^{\rho}p^{\sigma}+m_2k_1^{\rho}k_2^{\sigma}}{[p^2-m_2^2][(p+k_2)^2-m_1^2][(p-k_1)^2-m_2^2]}.
\eea

Using the Pauli-Villars regularization by introducing a set of fields with masses $M_1^j$, the subtracted tensor is
\bea \label{TTD2}
iq_{\rho}\widetilde{{M^{\pm}}}_{5,a}^{\rho\mu\nu}(k_1,k_2,m_1,m_2)=iq_{\rho}{M^{\pm}}_{5,a}^{\rho\mu\nu}(k_1,k_2,m_1,m_2)-i\sum_{j}c_j q_{\rho}{M^{\pm}}_{5,a}^{\rho\mu\nu}(k_1,k_2,M_1^j,\alpha M_1^j),
\eea
where the masses of the auxiliary fields are chosen to have the same ratio with that of physical fields, $\alpha=\frac{m_2}{m_1}$.
The divergences can be removed by imposing the conditions
\bea
\label{condition2}
\sum_{j}c_j=1, \quad \sum_{j}c_j (M_1^j)^2=m_1^2.
\eea
Taking the limit $M_1^j\rightarrow \infty$, the subtracted tensor in (\ref{TTD2}) becomes
\begin{align}
& iq_{\rho}\widetilde{{M^{\pm}}}_{5,a}^{\rho\mu\nu}(k_1,k_2,m_1,m_2)=\nonumber \\
&iN_c\frac{e_1g_WV_{12}(m_1+m_2)}{8\sqrt{2}\pi^2}\epsilon^{\mu\nu}_{\quad \rho\sigma}k_1^{\rho}k_2^{\sigma}\int_0^1 dx \int_0^{1-x}dy \frac{m_1(1-y)+ym_2}{(1-y)m_1^2+ym_2^2-x k_1^2-y k_2^2+(x k_1-y k_2)^2}\nonumber \\
&-iN_c\frac{e_1g_WV_{12}(m_1+m_2)}{8\sqrt{2}\pi^2}\epsilon^{\mu\nu}_{\quad \rho\sigma}k_1^{\rho}k_2^{\sigma}\int_0^1 dx \int_0^{1-x}dy \frac{m_1(1-y)+ym_2}{(1-y)m_1^2+ym_2^2}\nonumber \\
&+iN_c\frac{e_2g_WV_{12}(m_1+m_2)}{8\sqrt{2}\pi^2}\epsilon^{\mu\nu}_{\quad \rho\sigma}k_1^{\rho}k_2^{\sigma}\int_0^1 dx \int_0^{1-x}dy \frac{m_2(1-y)+ym_1}{(1-y)m_2^2+ym_1^2-x k_1^2-y k_2^2+(x k_1-y k_2)^2}\nonumber \\
&-iN_c\frac{e_2g_WV_{12}(m_1+m_2)}{8\sqrt{2}\pi^2}\epsilon^{\mu\nu}_{\quad \rho\sigma}k_1^{\rho}k_2^{\sigma}\int_0^1 dx \int_0^{1-x}dy \frac{m_2(1-y)+ym_1}{(1-y)m_2^2+ym_1^2}.
\end{align}
We again apply the integral formula (\ref{fz}) to compute $iq_{\rho}\widetilde{{M^{\pm}}}_{5,a}^{\rho\mu\nu}(k_1,k_2,m_1,m_2)$.
The procedure is the same as that in the previous section, the result for the integrals of $x$-variable for the first and the third terms is
\bea
&&\int_0^{1-x}dy \frac{m_1(1-y)+ym_2}{(1-y)m_1^2+ym_2^2-x k_1^2-y k_2^2+(x k_1-y k_2)^2}=\nonumber \\
&&P\int_0^{1-x}dy \frac{m_1(1-y)+ym_2}{(1-y)m_1^2+ym_2^2-x k_1^2-y k_2^2+(x k_1-y k_2)^2}+\nonumber \\
&&+\left[\frac{m_1(1-\hat{a})+\hat{a}m_2}{\hat{b}}-\frac{m_1-m_2}{2k_2^2}\right]2\pi in_1+\left[\frac{m_1(1-\hat{a})+\hat{a}m_2}{\hat{b}}-\frac{m_2-m_1}{2k_2^2}\right]2\pi in_2,
\eea
where the abbreviations $\hat{a}$ and $\hat{b}$ are defined by
\bea
&&\hat{a}=\frac{2xk_1\cdot k_2+k_2^2-(m_2^2-m_1^2)}{2k_2^2}, \nonumber\\
&& \hat{b}=\sqrt{\left[2xk_1\cdot k_2+k_2^2-(m_2^2-m_1^2)\right]^2-4(m_1^2-xk_1^2+x^2k_1^2)k_2^2}.
\eea
As before, it is possible to obtain the anomaly free condition $iq_{\rho}\widetilde{{M^{\pm}}}_{5,a}^{\rho\mu\nu}(k_1,k_2,m_1,m_2)=0$ in massless limit once the external momentums satisfy the following relation,
\bea
\label{aaa222}
&&\sqrt{(q^2-k_1^2-k_2^2)^2-4 k_1^2 k_2^2}\left[e_1I(m_1,m_2)+e_2I(m_2,m_1)\right]=\nonumber \\
&&(2\pi)^2ne_1\left[m_1(\frac{1}{2}+\frac{m_2^2-m_1^2}{2k_2^2})+m_2(\frac{1}{2}-\frac{m_2^2-m_1^2}{2k_2^2})\right]\nonumber \\
&&+(2\pi)^2ne_2\left[m_2(\frac{1}{2}+\frac{m_1^2-m_2^2}{2k_2^2})+m_1(\frac{1}{2}-\frac{m_1^2-m_2^2}{2k_2^2})\right],
\eea
where $n$ is an integer, $n\in \mathbb{Z}$, the expression $I(m_1,m_2)$ is
\begin{align}
I(m_1,m_2)=&\int_0^1 dx \int_0^{1-x}dy \frac{m_1(1-y)+ym_2}{(1-y)m_1^2+ym_2^2}\nonumber \\
=&\frac{m_1^4+2m_1^3 m_2-2m_1^2 m_2^2-2 m_1 m_2^3+m_2^4+2m_1m_2^3\log(m_2^2/m_1^2)}{2(m_1-m_2)^2(m_1+m_2)^3}.
\end{align}

We now consider the equation $ ik_{2\nu}{M^{\pm}}_{5}^{\rho\mu\nu}(k_1,k_2,m_1,m_2)\big{|}_{\epsilon}=0$. Dotting with $ik_{2\nu}$ yields
\begin{align} \label{aaa33}
&ik_{2\nu}{M^{\pm}}_{5,a}^{\rho\mu\nu}(k_1,k_2,m_1,m_2)=\nonumber\\
&iN_c\frac{e_1g_WV_{12}}{2\sqrt{2}}
\int\frac{d^4p}{(2\pi)^4}{\rm{Tr}}\left[\frac{p\!\!\!/+m_1}{p^2-m_1^2}\gamma^{\mu}\frac{p\!\!\!/+k\!\!\!/_1+m_1}{(p+k_1)^2-m_1^2}
\gamma^{\rho}\gamma^{5}\frac{p\!\!\!/-k\!\!\!/_2+m_2}{(p-k_2)^2-m_2^2}k\!\!\!/_2\right] \nonumber \\
&+iN_c\frac{e_2g_WV_{12}}{2\sqrt{2}}
\int\frac{d^4p}{(2\pi)^4}{\rm{Tr}}\left[\frac{p\!\!\!/+m_2}{p^2-m_2^2}k\!\!\!/_2\frac{p\!\!\!/+k\!\!\!/_2+m_1}{(p+k_2)^2-m_1^2}
\gamma^{\rho}\gamma^{5}\frac{p\!\!\!/-k\!\!\!/_1+m_2}{(p-k_1)^2-m_2^2}\gamma^{\mu}\right]=\nonumber \\
&-N_c\frac{e_1g_WV_{12}}{2\sqrt{2}}4(m_1-m_2)\epsilon^{\mu\rho}_{\quad \nu\sigma}\int\frac{d^4p}{(2\pi)^4}\frac{(m_2+m_1)p^{\nu}k_1^{\sigma}+m_1k_1^{\nu}k_2^{\sigma}}{[p^2-m_1^2][(p+k_1)^2-m_1^2][(p-k_2)^2-m_2^2]} \nonumber \\
&-N_c\frac{e_2g_WV_{12}}{2\sqrt{2}}4(m_1-m_2)\epsilon^{\mu\rho}_{\quad \nu\sigma}\int\frac{d^4p}{(2\pi)^4}\frac{(m_2+m_1)p^{\nu}k_1^{\sigma}-m_2k_1^{\nu}k_2^{\sigma}}{[p^2-m_2^2][(p+k_2)^2-m_1^2][(p-k_1)^2-m_2^2]}.
\end{align}
Without repeating the details of applying the integral formula (\ref{fz}) to compute, we give the anomaly free condition similar to (\ref{aaa222}),
\bea
\label{aaa22222}
&&\sqrt{(q^2-k_1^2-k_2^2)^2-4 k_1^2 k_2^2}\left(e_1I(m_1,-m_2)+e_2I(-m_2,m_1)\right)=\nonumber \\
&&(2\pi)^2ne_1\left[m_1(\frac{1}{2}+\frac{m_2^2-m_1^2}{2k_2^2})-m_2(\frac{1}{2}-\frac{m_2^2-m_1^2}{2k_2^2})\right]\nonumber \\
&&+(2\pi)^2ne_2\left[-m_2(\frac{1}{2}+\frac{m_1^2-m_2^2}{2k_2^2})+m_1(\frac{1}{2}-\frac{m_1^2-m_2^2}{2k_2^2})\right].
\eea
The masses of mesons $\pi^{\pm}$ and $K^{\pm}$ are related to the on-shell relation $q^2=m_P^2$, they can be obtained by solving the equations (\ref{aaa222}) and (\ref{aaa22222}).

\section{The mass formula of mesons $K^0$ and $\bar{K}^0$}\label{MassesKandKBAR}

\begin{figure}[H]
\center
\includegraphics[width=0.5\textwidth]{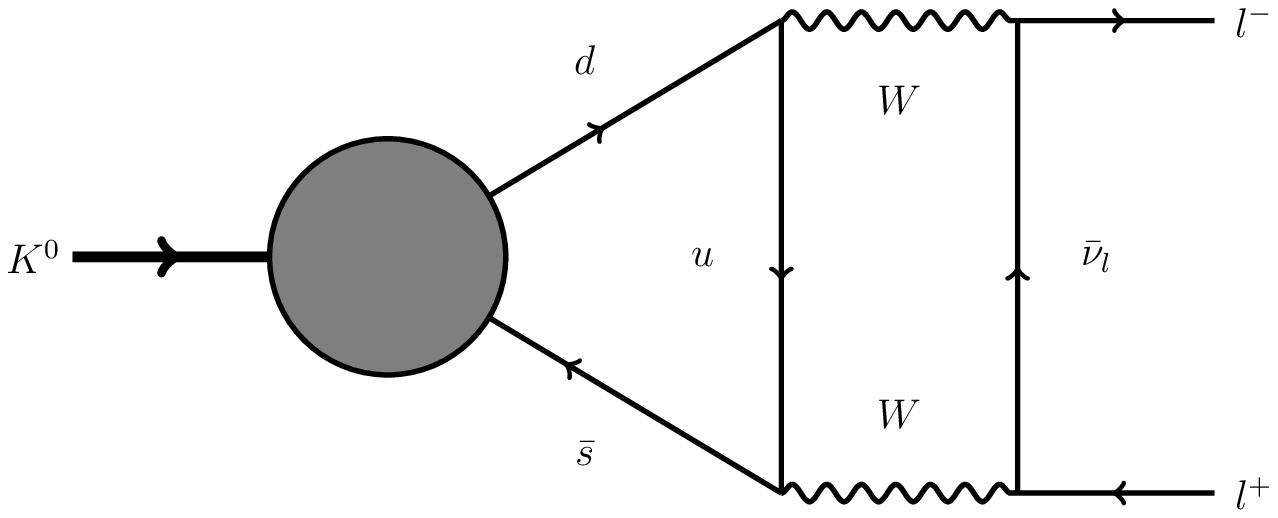}
\caption{\label{K0decay} The decay channel of $K^0\to l^-l^+$. }
\end{figure}

\begin{figure}[H]
\center
\includegraphics[width=0.6\textwidth]{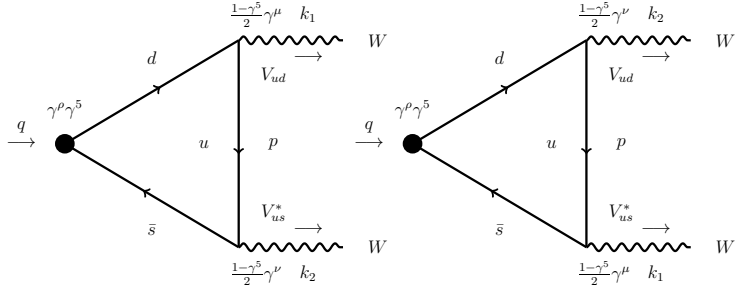}
\caption{\label{triangle3} Triangle diagrams contributing to calculate the divergence of current $j^{\rho5}=i\bar{s}\gamma^{\rho}\gamma^5 d$.  }
\end{figure}

Similar to the case of $\pi$-meson, the $K$-meson has decay channels to leptons as shown in Figure \ref{K0decay}.
The diagram contains a subdiagram where the ABJ anomaly might arise, as shown in Figure \ref{triangle3}.
The axial vector currents corresponding to the mesons $K^0$ and $\bar{K}^0$ have the form $j^{\rho5}=i\bar{\psi}_1\gamma^{\rho}\gamma^5 \psi_2$, with $\psi_1, \psi_2=d, s$. We calculate the triangle diagrams, in which both $k_1$ and $k_2$ are off-shell momenta of intermediate heavy gauge bosons, and each vertex with $W$-boson attaches a chirality projection operator $(1-\gamma_5)/2$, the corresponding tensor is
\bea\label{section4Mtensor}
&&{M^{0}}_{5}^{\rho\mu\nu}(k_1,k_2,m_u,m_d,m_s)=\nonumber \\
&&N_c(\frac{g_W}{2\sqrt{2}})^2V_{ud}{V^{\ast}}_{u\bar{s}}
\int\frac{d^4p}{(2\pi)^4}{\rm{Tr}}\left[\frac{i}{p\!\!\!/-m_u}(1-\gamma^5)\gamma^{\mu}\frac{i}{p\!\!\!/+k\!\!\!/_1-m_d}
i\gamma^{\rho}\gamma^{5}\frac{i}{p\!\!\!/-k\!\!\!/_2-m_s}(1-\gamma^5)\gamma^{\nu}\right] \nonumber \\
&&+\left(
  \begin{array}{c}
    \mu\leftrightarrow\nu \\
    k_1\leftrightarrow k_2 \\
  \end{array}
\right).
\eea

In massless limit, the anomaly free conditions are the following,
\bea
\label{qm1}
\begin{cases}
  iq_{\rho}{M^{0}}_{5}^{\rho\mu\nu}(k_1,k_2,m_u,m_d,m_s)\big{|}_{\epsilon}=0, \\
   ik_{1\mu}{M^{0}}_{5}^{\rho\mu\nu}(k_1,k_2,m_u,m_d,m_s)\big{|}_{\epsilon}=0, \\
   ik_{2\nu}{M^{0}}_{5}^{\rho\mu\nu}(k_1,k_2,m_u,m_d,m_s)\big{|}_{\epsilon}=0,
 \end{cases}
\eea
where only terms containing the Levi-Civita symbol are needed.
To compute the tensor in the equations (\ref{qm1}), where there are three $\gamma_5$ matrices,
one encounter an problem of manipulating the order of gamma matrices in dimensional regularization: there is no well-defined rule to order the matrix $\gamma_5$ \cite{Jegerlehner:2000dz}. At the moment there are different approaches to this problem which  lead to different results on some anomaly related quantities, e. g. the trace anomaly \cite{Bonora:2014qla,Bonora:2017gzz,Bonora:2020upc,Liu:2022jxz,Liu:2023yhh,Bonora:2022izj,Bastianelli:2016nuf,Abdallah:2021eii,Bastianelli:2022hmu}.

One way is to move one projection operator $(1-\gamma^5)/2$ without changing its form and combine with another projection operator into a single factor without, that means the following,
\bea
\cdots\left(\frac{1-\gamma^5}{2}\right)\gamma^{\nu}\frac{i}{p\!\!\!/-m_u}\left(\frac{1-\gamma^5}{2}\right)\gamma^{\mu}\cdots
=\cdots\gamma^{\nu}\frac{i}{p\!\!\!/-m_u}\left(\frac{1-\gamma^5}{2}\right)\gamma^{\mu}\cdots.
\eea
We call this prescription as single projector prescription.
This method has been used in calculating the trace anomaly \cite{Bonora:2014qla,Bonora:2017gzz,Liu:2023yhh}.

The second way is to use the definition $\gamma^5=i\gamma^0\gamma^1\gamma^2\gamma^3$, substitution of $\gamma_5$ by the relation leads to
\bea
\cdots\left(\frac{1-\gamma^5}{2}\right)\gamma^{\nu}\frac{i}{p\!\!\!/-m_u}\left(\frac{1-\gamma^5}{2}\right)\gamma^{\mu}\cdots
=\cdots\gamma^{\nu}\frac{ip\!\!\!/}{p^2-m_u^2}\left(\frac{1-\gamma^5}{2}\right)\gamma^{\mu}\cdots.
\eea
This is similar with the Breitenlohner-Maison-`t Hooft-Veltman prescription \cite{tHooft:1972tcz,Breitenlohner:1977hr,Breitenlohner:1975hg,Breitenlohner:1976te}
in dimensional regularization. In dimensional regularization, this definition has consequence that
\bea
\begin{cases}
   \{\gamma^5, \gamma^{\mu}\}=0 & {\rm{for}}\:\: \mu=0,1,2,3; \\
    [\gamma^5, \gamma^{\mu}]=0 &{\rm{for}}\:\: \mu=4,\ldots,d-1.
 \end{cases}
\eea
These prescription is used in the calculation of the trace anomaly in the works \cite{Abdallah:2021eii,Bastianelli:2022hmu}. For convenience, we call  the second way as the normal Pauli-Villars prescription. In the remaining of this section, both prescriptions are used to calculate the tensor ${M^{0}}_{5}^{\rho\mu\nu}(k_1,k_2,m_u,m_d,m_s)$.

\subsection{Single projector prescription}

Simplifying the gamma matrix factors involving $\gamma_5$ in (\ref{section4Mtensor}) using the single projector prescription, ${M^{0}}_{5}^{\rho\mu\nu}(k_1,k_2,m_u,m_d,m_s)$ becomes
\bea
&&{M^{0}}_{5}^{\rho\mu\nu}(k_1,k_2,m_u,m_d,m_s)=\nonumber \\
&&N_c(\frac{g_W}{2})^2V_{ud}{V^{\ast}}_{u\bar{s}}
\int\frac{d^4p}{(2\pi)^4}{\rm{Tr}}\left[\frac{i}{p\!\!\!/-m_u}\gamma^{\mu}\frac{i}{p\!\!\!/+k\!\!\!/_1-m_d}
i\gamma^{\rho}\gamma^{5}\frac{i}{p\!\!\!/-k\!\!\!/_2-m_s}(1-\gamma^5)\gamma^{\nu}\right] \nonumber \\
&&+\left(
  \begin{array}{c}
    \mu\leftrightarrow\nu \\
    k_1\leftrightarrow k_2 \\
  \end{array}
\right).
\eea
The $\epsilon$-dependent piece comes from the anomaly term of ${M^{0}}_{5,a}^{\rho\mu\nu}(k_1,k_2,m_u,m_d,m_s)$ which is calculated by
\bea
&&{M^{0}}_{5,a}^{\rho\mu\nu}(k_1,k_2,m_u,m_d,m_s)=\nonumber \\
&&N_c(\frac{g_W}{2})^2V_{ud}{V^{\ast}}_{u\bar{s}}
\int\frac{d^4p}{(2\pi)^4}{\rm{Tr}}\left[\frac{i}{p\!\!\!/-m_u}\gamma^{\mu}\frac{i}{p\!\!\!/+k\!\!\!/_1-m_d}
i\gamma^{\rho}\gamma^{5}\frac{i}{p\!\!\!/-k\!\!\!/_2-m_s}\gamma^{\nu}\right] \nonumber \\
&&+\left(
  \begin{array}{c}
    \mu\leftrightarrow\nu \\
    k_1\leftrightarrow k_2 \\
  \end{array}
\right).
\eea
Dotting with $iq^{\rho}$ yields
\bea \label{aaa2}
&&iq_{\rho}{M^{0}}_{5,a}^{\rho\mu\nu}(k_1,k_2,m_u,m_d,m_s)=\nonumber \\
&&iN_c(\frac{g_W}{2})^2V_{ud}{V^{\ast}}_{u\bar{s}}
\int\frac{d^4p}{(2\pi)^4}{\rm{Tr}}\left[\frac{p\!\!\!/+m_u}{p^2-m_u^2}\gamma^{\mu}\frac{p\!\!\!/+k\!\!\!/_1+m_d}{(p+k_1)^2-m_d^2}
q\!\!\!/\gamma^{5}\frac{p\!\!\!/-k\!\!\!/_2+m_s}{(p-k_2)^2-m_s^2}\gamma^{\nu}\right] \nonumber \\
&&+iN_c(\frac{g_W}{2})^2V_{ud}{V^{\ast}}_{u\bar{s}}
\int\frac{d^4p}{(2\pi)^4}{\rm{Tr}}\left[\frac{p\!\!\!/+m_u}{p^2-m_u^2}\gamma^{\nu}\frac{p\!\!\!/+k\!\!\!/_2+m_d}{(p+k_2)^2-m_d^2}
q\!\!\!/\gamma^{5}\frac{p\!\!\!/-k\!\!\!/_1+m_s}{(p-k_1)^2-m_s^2}\gamma^{\mu}\right]
\nonumber \\
&&=iN_c(\frac{g_W}{4\pi})^2V_{ud}{V^{\ast}}_{u\bar{s}}(m_d+m_s)
\epsilon^{\mu\nu}_{\quad \rho\sigma}\nonumber \\
 &&\times\int_0^1 dx \int_0^{1-x}dy\frac{ \left[x (m_d-m_u)+y( m_s-m_u)+m_u\right]k_1^{\rho} k_2^{\sigma}}{x m_d^2+ym_s^2+(1-x-y)m_u^2-x k_1^2-y k_2^2+(x k_1-y k_2)^2}\nonumber \\
&&+\left(
  \begin{array}{c}
    \mu\leftrightarrow\nu \\
    k_1\leftrightarrow k_2 \\
  \end{array}
\right).
\eea
In Pauli-Villars regularization, we use a set of fields \cite{Pauli:1949zm} with masses $M_u^j$, the subtracted tensor is computed by
\bea \label{TTD4}
&&iq_{\rho}\widetilde{{M^{0}}}_{5,a}^{\rho\mu\nu}(k_1,k_2,m_u,m_d,m_s)=\nonumber\\
&&iq_{\rho}{M^{0}}_{5,a}^{\rho\mu\nu}(k_1,k_2,m_u,m_d,m_s)
-i\sum_{j}c_j q_{\rho}{M^{0}}_{5,a}^{\rho\mu\nu}(k_1,k_2,M_u^j,\beta M_u^j,\chi M_u^j),
\eea
where the constants $\beta=\frac{m_d}{m_u}$ and $\chi=\frac{m_s}{m_u}$ are chosen to ensure the ratios of regulator fields are the same as that of the quark fields. To remove the divergences, the following conditions are imposed,
\bea
\label{condition4}
\sum_{j}c_j=1, \quad \sum_{j}c_j (M_u^j)^2=m_u^2.
\eea
Then taking the limit $M_u^j\rightarrow \infty$, (\ref{TTD4}) becomes
\bea
&&iq_{\rho}\widetilde{{M^{0}}}_{5,a}^{\rho\mu\nu}(k_1,k_2,m_u,m_d,m_s)=\nonumber \\
&&iN_c(\frac{g_W}{4\pi})^2V_{ud}{V^{\ast}}_{u\bar{s}}(m_d+m_s)
\epsilon^{\mu\nu}_{\quad \rho\sigma} \nonumber \\
&&\times\int_0^1 dx \int_0^{1-x}dy\frac{ \left[x (m_d-m_u)+y( m_s-m_u)+m_u\right]k_1^{\rho} k_2^{\sigma}}{x m_d^2+ym_s^2+(1-x-y)m_u^2-x k_1^2-y k_2^2+(x k_1-y k_2)^2}\nonumber \\
&&-iN_c(\frac{g_W}{4\pi})^2V_{ud}{V^{\ast}}_{u\bar{s}}(m_d+m_s)
\epsilon^{\mu\nu}_{\quad \rho\sigma}
\int_0^1 dx \int_0^{1-x}dy\frac{ \left[x (m_d-m_u)+y( m_s-m_u)+m_u\right]k_1^{\rho} k_2^{\sigma}}{x m_d^2+ym_s^2+(1-x-y)m_u^2}\nonumber \\
&&+\left(
  \begin{array}{c}
    \mu\leftrightarrow\nu \\
    k_1\leftrightarrow k_2 \\
  \end{array}
\right).
\eea
As in previous examples, we then apply the integral formula (\ref{fz}) to calculate  $iq_{\rho}\widetilde{{M^{0}}}_{5,a}^{\rho\mu\nu}(k_1,k_2,m_u,m_d,m_s)$.
The external momentums $k_1^2$ and $k_2^2$ are associated with massive gauge bosons, therefore they satisfy $k_1^2\neq 0$ and $k_2^2\neq 0$.
The procedure leads to the following result,
\bea
&&\int_0^{1-x}dy\frac{ \left[x (m_d-m_u)+y( m_s-m_u)+m_u\right]}{x m_d^2+ym_s^2+(1-x-y)m_u^2-x k_1^2-y k_2^2+(x k_1-y k_2)^2} \nonumber \\
&&=P \int_0^{1-x}dy\frac{ \left[x (m_d-m_u)+y( m_s-m_u)+m_u\right]}{x m_d^2+ym_s^2+(1-x-y)m_u^2-x k_1^2-y k_2^2+(x k_1-y k_2)^2} \nonumber \\
&&+\frac{2 \pi i q}{k_2^2 \sqrt{c^2-4 d}}\left[\frac{-c+\sqrt{c^2-4 d}}{2}(m_s-m_u)+m_u+x(m_d-m_u)\right]+\frac{2 \pi i l}{k_2^2}(m_s-m_u),
\eea
where $l$ is an integer, $l\in\mathbb{Z}$, the abbreviations $c$ and $d$ are defined by
\bea
\label{cd}
c=\frac{m_s^2-m_u^2-k_2^2-2xk_1\cdot k_2}{k_2^2}, \quad d=\frac{xm_d^2+(1-x)m_u^2-xk_1^2+x^2 k_1^2}{k_2^2}.
\eea
The anomaly free condition $iq_{\rho}\widetilde{{M^{0}}}_{5,a}^{\rho\mu\nu}(k_1,k_2,m_u,m_d,m_s)=0$ in the massless limit is satisfied when the external momentums satisfy the following relations,
\bea
\label{aaa31}
&&\frac{(2\pi)^2n}{\sqrt{(q^2-k_1^2-k_2^2)^2-4 k_1^2 k_2^2}}\left[m_s(\frac{1}{2}-\frac{m_s^2-m_u^2}{2k_2^2})+m_u(\frac{1}{2}+\frac{m_s^2-m_u^2}{2k_2^2})\right.\nonumber \\
&&\left.+m_s(\frac{1}{2}-\frac{m_s^2-m_u^2}{2k_1^2})+m_u(\frac{1}{2}+\frac{m_s^2-m_u^2}{2k_1^2})\right]=2J(m_u,m_d,m_s),
\eea
where the $J(m_u,m_d,m_s)$ is defined by
\bea
\label{J}
J(m_u,m_d,m_s)=\int_0^1 dx \int_0^{1-x}dy\frac{ x (m_d-m_u)+y( m_s-m_u)+m_u}{x m_d^2+ym_s^2+(1-x-y)m_u^2}.
\eea

The others equations in (\ref{qm1}) can be treated with the same method, here we give the results as the following.
For the second condition in (\ref{qm1}), dotting with $ik_{1\mu}$ yields
\bea
&&ik_{1\mu}{M^{0}}_{5,a}^{\rho\mu\nu}(k_1,k_2,m_u,m_d,m_s)=\nonumber \\
&&N_c(\frac{g_W}{2})^2V_{ud}{V^{\ast}}_{u\bar{s}}
\int\frac{d^4p}{(2\pi)^4}{\rm{Tr}}\left[\frac{i}{p\!\!\!/-m_u}ik\!\!\!/_1\frac{i}{p\!\!\!/+k\!\!\!/_1-m_d}
i\gamma^{\rho}\gamma^{5}\frac{i}{p\!\!\!/-k\!\!\!/_2-m_s}\gamma^{\nu}\right] \nonumber \\
&&+N_c(\frac{g_W}{2})^2V_{ud}{V^{\ast}}_{u\bar{s}}
\int\frac{d^4p}{(2\pi)^4}{\rm{Tr}}\left[\frac{i}{p\!\!\!/-m_u}\gamma^{\nu}\frac{i}{p\!\!\!/+k\!\!\!/_2-m_d}
i\gamma^{\rho}\gamma^{5}\frac{i}{p\!\!\!/-k\!\!\!/_1-m_s}ik\!\!\!/_1\right]  \\
&&=iN_c(\frac{g_W}{2})^2V_{ud}{V^{\ast}}_{u\bar{s}}
(m_d-m_u)\epsilon^{\nu\rho}_{\quad \mu\sigma}\int\frac{d^4p}{(2\pi)^4}\frac{4i(m_u-m_s) p^{\mu} k_1^{\sigma}+4i(m_u+m_d) p^{\mu}k_2^{\sigma}+4im_u k_1^{\mu} k_2^{\sigma}}{[p^2-m_u^2][(p+k_1)^2-m_d^2][(p-k_2)^2-m_s^2]}\nonumber\\
&&+iN_c(\frac{g_W}{2})^2V_{ud}{V^{\ast}}_{u\bar{s}}
(m_u-m_s)\epsilon^{\nu\rho}_{\quad \mu\sigma}\int\frac{d^4p}{(2\pi)^4}\frac{4i(m_u-m_d) p^{\mu} k_1^{\sigma}+4i(m_u+m_s) p^{\mu}k_2^{\sigma}-4im_u k_1^{\mu} k_2^{\sigma}}{[p^2-m_u^2][(p+k_2)^2-m_d^2][(p-k_1)^2-m_s^2]}.\nonumber
\eea
Using the same method to perform integration, we obtain
\bea
\label{aaa32}
&&\frac{(2\pi)^2n}{\sqrt{(q^2-k_1^2-k_2^2)^2-4 k_1^2 k_2^2}}\left[(m_d-m_u)\left(m_s(\frac{1}{2}-\frac{m_s^2-m_u^2}{2k_2^2})+m_u(\frac{1}{2}+\frac{m_s^2-m_u^2}{2k_2^2})\right)\right.\nonumber \\
&&\left.+(m_u-m_s)\left(m_s(\frac{1}{2}-\frac{m_s^2-m_u^2}{2k_1^2})-m_u(\frac{1}{2}+\frac{m_s^2-m_u^2}{2k_1^2})\right)\right]\nonumber \\
&&=(m_d-m_u)J(m_u,-m_d,m_s)+(m_u-m_s)J(-m_u,-m_d,m_s).
\eea
For the third condition in (\ref{qm1}), dotting with $ik_{2\nu}$ yields
\bea
&&ik_{2\nu}{M^{0}}_{5,a}^{\rho\mu\nu}(k_1,k_2,m_u,m_d,m_s)=\nonumber \\
&&N_c(\frac{g_W}{2})^2V_{ud}{V^{\ast}}_{u\bar{s}}
\int\frac{d^4p}{(2\pi)^4}{\rm{Tr}}\left[\frac{i}{p\!\!\!/-m_u}\gamma^{\mu}\frac{i}{p\!\!\!/+k\!\!\!/_1-m_d}
i\gamma^{\rho}\gamma^{5}\frac{i}{p\!\!\!/-k\!\!\!/_2-m_s}ik\!\!\!/_2\right] \nonumber \\
&&+N_c(\frac{g_W}{2})^2V_{ud}{V^{\ast}}_{u\bar{s}}
\int\frac{d^4p}{(2\pi)^4}{\rm{Tr}}\left[\frac{i}{p\!\!\!/-m_u}ik\!\!\!/_2\frac{i}{p\!\!\!/+k\!\!\!/_2-m_d}
i\gamma^{\rho}\gamma^{5}\frac{i}{p\!\!\!/-k\!\!\!/_1-m_s}\gamma^{\mu}\right].
\eea
The result of performing integration is
\bea
\label{aaa33}
&&\frac{(2\pi)^2n}{\sqrt{(q^2-k_1^2-k_2^2)^2-4 k_1^2 k_2^2}}\left[(m_d-m_u)\left(m_s(\frac{1}{2}-\frac{m_s^2-m_u^2}{2k_1^2})+m_u(\frac{1}{2}+\frac{m_s^2-m_u^2}{2k_1^2})\right)\right.\nonumber \\
&&\left.+(m_u-m_s)\left(m_s(\frac{1}{2}-\frac{m_s^2-m_u^2}{2k_2^2})-m_u(\frac{1}{2}+\frac{m_s^2-m_u^2}{2k_2^2})\right)\right]\nonumber \\
&&=(m_d-m_u)J(m_u,-m_d,m_s)+(m_u-m_s)J(-m_u,-m_d,m_s).
\eea
From equations (\ref{aaa32}) and (\ref{aaa33}), it is easy to find the condition $k_1^2=k_2^2$, it is reasonable since they both are related to on-shell W-boson mass. The masses of $K^0$ and $\bar{K}^0$ can be obtained by solving the equations (\ref{aaa31}), (\ref{aaa32}) and (\ref{aaa33}).

\subsection{Normal Pauli-Villars prescription}
We now use normal Pauli-Villars prescription to solve the equations in (\ref{qm1}). Dotting with $iq^{\rho}$ yields
\bea \label{aaa2}
&&iq_{\rho}{M^{0}}_{5}^{\rho\mu\nu}(k_1,k_2,m_u,m_d,m_s)=\nonumber \\
&&iN_c(\frac{g_W}{2\sqrt{2}})^2V_{ud}{V^{\ast}}_{u\bar{s}}
\int\frac{d^4p}{(2\pi)^4}{\rm{Tr}}\left[\frac{p\!\!\!/+m_u}{p^2-m_u^2}(1-\gamma^5)\gamma^{\mu}\frac{p\!\!\!/+k\!\!\!/_1+m_d}{(p+k_1)^2-m_d^2}
q\!\!\!/\gamma^{5}\frac{p\!\!\!/-k\!\!\!/_2+m_s}{(p-k_2)^2-m_s^2}(1-\gamma^5)\gamma^{\nu}\right] \nonumber \\
&&+\left(
  \begin{array}{c}
    \mu\leftrightarrow\nu \\
    k_1\leftrightarrow k_2 \\
  \end{array}
\right).
\eea
The $\epsilon$-dependent piece is
\bea
&&iq_{\rho}{M^{0}}_{5}^{\rho\mu\nu}(k_1,k_2,m_u,m_d,m_s)\bigg{|}_{\epsilon}=\nonumber \\
&&iN_c(\frac{g_W}{2\sqrt{2}})^2V_{ud}{V^{\ast}}_{u\bar{s}}(m_d+m_s)
\epsilon^{\mu\nu}_{\quad \rho\sigma}\int\frac{d^4p}{(2\pi)^4}\frac{-8im_s p^{\rho} k_1^{\sigma}-8im_d p^{\rho} k_2^{\sigma}}{[p^2-m_u^2][(p+k_1)^2-m_d^2][(p-k_2)^2-m_s^2]}\nonumber \\
&&+\left(
  \begin{array}{c}
    \mu\leftrightarrow\nu \\
    k_1\leftrightarrow k_2 \\
  \end{array}
\right)
\nonumber \\
&&=iN_c(\frac{g_W}{4\pi})^2V_{ud}{V^{\ast}}_{u\bar{s}}(m_d+m_s)
\epsilon^{\mu\nu}_{\quad \rho\sigma}\nonumber \\
&&\times \int_0^1 dx \int_0^{1-x}dy\frac{ (x m_d+y m_s)k_1^{\rho} k_2^{\sigma}}{x m_d^2+ym_s^2+(1-x-y)m_u^2-x k_1^2-y k_2^2+(x k_1-y k_2)^2}\nonumber \\
&&+\left(
  \begin{array}{c}
    \mu\leftrightarrow\nu \\
    k_1\leftrightarrow k_2 \\
  \end{array}
\right).
\eea
To regulate the integral one introduces a set of Pauli-Villars fields \cite{Pauli:1949zm} with masses $M_u^j$, then the subtracted tensor is
\bea \label{TTD3}
&&iq_{\rho}\widetilde{{M^{0}}}_{5}^{\rho\mu\nu}(k_1,k_2,m_u,m_d,m_s)\bigg{|}_{\epsilon}=\nonumber\\
&&iq_{\rho}{M^{0}}_{5}^{\rho\mu\nu}(k_1,k_2,m_u,m_d,m_s)\bigg{|}_{\epsilon}
-i\sum_{j}c_j q_{\rho}{M^{0}}_{5}^{\rho\mu\nu}(k_1,k_2,M_u^j,\beta M_u^j,\chi M_u^j)\bigg{|}_{\epsilon},
\eea
where $\beta=\frac{m_d}{m_u}$ and $\chi=\frac{m_s}{m_u}$ are constants introduced for the same reason as before.
The conditions $\sum_{j}c_j=1$, $\sum_{j}c_j (M_u^j)^2=m_u^2$ are imposed to remove the divergences.
Taking the limit $M_u^j\rightarrow \infty$, (\ref{TTD3}) becomes
\bea
&&iq_{\rho}\widetilde{{M^{0}}}_{5}^{\rho\mu\nu}(k_1,k_2,m_u,m_d,m_s)\bigg{|}_{\epsilon}=\nonumber \\
&&iN_c(\frac{g_W}{4\pi})^2V_{ud}{V^{\ast}}_{u\bar{s}}(m_d+m_s)
\epsilon^{\mu\nu}_{\quad \rho\sigma}\nonumber \\
&&\times \int_0^1 dx \int_0^{1-x}dy\frac{ (x m_d+y m_s)k_1^{\rho} k_2^{\sigma}}{x m_d^2+ym_s^2+(1-x-y)m_u^2-x k_1^2-y k_2^2+(x k_1-y k_2)^2}\nonumber \\
&&-iN_c(\frac{g_W}{4\pi})^2V_{ud}{V^{\ast}}_{u\bar{s}}(m_d+m_s)
\epsilon^{\mu\nu}_{\quad \rho\sigma} \int_0^1 dx \int_0^{1-x}dy\frac{ (x m_d+y m_s)k_1^{\rho} k_2^{\sigma}}{x m_d^2+ym_s^2+(1-x-y)m_u^2}\nonumber \\
&&+\left(
  \begin{array}{c}
    \mu\leftrightarrow\nu \\
    k_1\leftrightarrow k_2 \\
  \end{array}
\right).
\eea
We then apply the integral formula (\ref{fz}) to study $iq_{\rho}\widetilde{{M^{0}}}_{5}^{\rho\mu\nu}(k_1,k_2,m_u,m_d,m_s)\bigg{|}_{\epsilon}$. Calculating the integral, we obtain
\bea
&&\int_0^{1-x}dy\frac{ \left(x m_d+ym_s\right)}{x m_d^2+ym_s^2+(1-x-y)m_u^2-x k_1^2-y k_2^2+(x k_1-y k_2)^2} \nonumber \\
&&=P \int_0^{1-x}dy\frac{ \left(x m_d+y m_s\right)}{x m_d^2+ym_s^2+(1-x-y)m_u^2-x k_1^2-y k_2^2+(x k_1-y k_2)^2} \nonumber \\
&&+\frac{2 \pi i q}{k_2^2 \sqrt{c^2-4 d}}\left[\frac{-c+\sqrt{c^2-4 d}}{2}m_s+xm_d\right]+\frac{2 \pi i l}{k_2^2}m_s,
\eea
where $c$ and $d$ are given by the same expressions in (\ref{cd}).
To satisfy the anomaly free condition $iq_{\rho}\widetilde{{M^{0}}}_{5,a}^{\rho\mu\nu}(k_1,k_2,m_u,m_d,m_s)=0$ in the massless limit, the external momentums should satisfy the relation
\bea
\label{bbb1}
\frac{(2\pi)^2n}{\sqrt{(q^2-k_1^2-k_2^2)^2-4 k_1^2 k_2^2}}\left[m_s(\frac{1}{2}-\frac{m_s^2-m_u^2}{2k_2^2})
+m_s(\frac{1}{2}-\frac{m_s^2-m_u^2}{2k_1^2})\right]=2K(m_u,m_d,m_s),
\eea
where the $K(m_u,m_d,m_s)$ is defined by
\bea
K(m_u,m_d,m_s)=\int_0^1 dx \int_0^{1-x}dy\frac{ (x m_d+y m_s)}{x m_d^2+ym_s^2+(1-x-y)m_u^2}.
\eea

The other equations in (\ref{qm1}) can be studied following the same method.
Dotting with $ik_{1\mu}$ yields
\bea
&&ik_{1\mu}{M^{0}}_{5}^{\rho\mu\nu}(k_1,k_2,m_u,m_d,m_s)=\nonumber \\
&&N_c(\frac{g_W}{2\sqrt{2}})^2V_{ud}{V^{\ast}}_{u\bar{s}}
\int\frac{d^4p}{(2\pi)^4}{\rm{Tr}}\left[\frac{i}{p\!\!\!/-m_u}(1-\gamma^5)ik\!\!\!/_1\frac{i}{p\!\!\!/+k\!\!\!/_1-m_d}
i\gamma^{\rho}\gamma^{5}\frac{i}{p\!\!\!/-k\!\!\!/_2-m_s}(1-\gamma^5)\gamma^{\nu}\right] \nonumber \\
&&+N_c(\frac{g_W}{2\sqrt{2}})^2V_{ud}{V^{\ast}}_{u\bar{s}}
\int\frac{d^4p}{(2\pi)^4}{\rm{Tr}}\left[\frac{i}{p\!\!\!/-m_u}(1-\gamma^5)\gamma^{\nu}\frac{i}{p\!\!\!/+k\!\!\!/_2-m_d}
i\gamma^{\rho}\gamma^{5}\frac{i}{p\!\!\!/-k\!\!\!/_1-m_s}(1-\gamma^5)ik\!\!\!/_1\right].
\eea
The $\epsilon$-dependent piece is
\bea
&&ik_{1\mu}{M^{0}}_{5}^{\rho\mu\nu}(k_1,k_2,m_u,m_d,m_s)\bigg{|}_{\epsilon}=\nonumber \\
&&iN_c(\frac{g_W}{2\sqrt{2}})^2V_{ud}{V^{\ast}}_{u\bar{s}}
(m_d-m_u)\epsilon^{\nu\rho}_{\quad \mu\sigma}\int\frac{d^4p}{(2\pi)^4}\frac{4i(m_u-m_s) p^{\mu} k_1^{\sigma}+4i(m_u+m_d) p^{\mu}k_2^{\sigma}+4im_u k_1^{\mu} k_2^{\sigma}}{[p^2-m_u^2][(p+k_1)^2-m_d^2][(p-k_2)^2-m_s^2]}\nonumber\\
&&-iN_c(\frac{g_W}{2\sqrt{2}})^2V_{ud}{V^{\ast}}_{u\bar{s}}
(m_d+m_u)\epsilon^{\nu\rho}_{\quad \mu\sigma}\int\frac{d^4p}{(2\pi)^4}\frac{4i(m_u+m_s) p^{\mu} k_1^{\sigma}+4i(m_u-m_d) p^{\mu}k_2^{\sigma}+4im_u k_1^{\mu} k_2^{\sigma}}{[p^2-m_u^2][(p+k_1)^2-m_d^2][(p-k_2)^2-m_s^2]}\nonumber\\
&&+iN_c(\frac{g_W}{2\sqrt{2}})^2V_{ud}{V^{\ast}}_{u\bar{s}}
(m_u-m_s)\epsilon^{\nu\rho}_{\quad \mu\sigma}\int\frac{d^4p}{(2\pi)^4}\frac{4i(m_u-m_d) p^{\mu} k_1^{\sigma}+4i(m_u+m_s) p^{\mu}k_2^{\sigma}-4im_u k_1^{\mu} k_2^{\sigma}}{[p^2-m_u^2][(p+k_2)^2-m_d^2][(p-k_1)^2-m_s^2]}\nonumber\\
&&-iN_c(\frac{g_W}{2\sqrt{2}})^2V_{ud}{V^{\ast}}_{u\bar{s}}
(m_u+m_s)\epsilon^{\nu\rho}_{\quad \mu\sigma}\int\frac{d^4p}{(2\pi)^4}\frac{4i(-m_u-m_d) p^{\mu} k_1^{\sigma}+4i(-m_u+m_s) p^{\mu}k_2^{\sigma}+4im_u k_1^{\mu} k_2^{\sigma}}{[p^2-m_u^2][(p+k_2)^2-m_d^2][(p-k_1)^2-m_s^2]}.
\eea
Application of the integral formula (\ref{fz}) leads to the result,
\bea
\label{bbb2}
&&\frac{(2\pi)^2n}{\sqrt{(q^2-k_1^2-k_2^2)^2-4 k_1^2 k_2^2}}\left[(m_d-m_u)\left(m_s(\frac{1}{2}-\frac{m_s^2-m_u^2}{2k_2^2})+m_u(\frac{1}{2}+\frac{m_s^2-m_u^2}{2k_2^2})\right)\right.\nonumber \\
&&\left.-(m_d+m_u)\left(-m_s(\frac{1}{2}-\frac{m_s^2-m_u^2}{2k_2^2})+m_u(\frac{1}{2}+\frac{m_s^2-m_u^2}{2k_2^2})\right)\right.\nonumber \\
&&\left.+(m_u-m_s)\left(m_s(\frac{1}{2}-\frac{m_s^2-m_u^2}{2k_1^2})-m_u(\frac{1}{2}+\frac{m_s^2-m_u^2}{2k_1^2})\right)\right.\nonumber \\
&&\left.-(m_u+m_s)\left(m_s(\frac{1}{2}-\frac{m_s^2-m_u^2}{2k_1^2})+m_u(\frac{1}{2}+\frac{m_s^2-m_u^2}{2k_1^2})\right)\right]\nonumber \\
&&=(m_d-m_u)J(m_u,-m_d,m_s)-(m_d+m_u)J(m_u,m_d,-m_s)\nonumber \\
&&+(m_u-m_s)J(-m_u,-m_d,m_s)-(m_u+m_s)J(m_u,-m_d,m_s),
\eea
where $J(m_u,m_d,m_s)$ is the same function given by (\ref{J}). Likewise, dotting with $ik_{2\nu}$ yields
\bea
&&ik_{2\nu}{M^{0}}_{5}^{\rho\mu\nu}(k_1,k_2,m_u,m_d,m_s)=\nonumber \\
&&N_c(\frac{g_W}{2\sqrt{2}})^2V_{ud}{V^{\ast}}_{u\bar{s}}
\int\frac{d^4p}{(2\pi)^4}{\rm{Tr}}\left[\frac{i}{p\!\!\!/-m_u}(1-\gamma^5)\gamma^{\mu}\frac{i}{p\!\!\!/+k\!\!\!/_1-m_d}
i\gamma^{\rho}\gamma^{5}\frac{i}{p\!\!\!/-k\!\!\!/_2-m_s}(1-\gamma^5)ik\!\!\!/_2\right] \nonumber \\
&&+N_c(\frac{g_W}{2\sqrt{2}})^2V_{ud}{V^{\ast}}_{u\bar{s}}
\int\frac{d^4p}{(2\pi)^4}{\rm{Tr}}\left[\frac{i}{p\!\!\!/-m_u}(1-\gamma^5)ik\!\!\!/_2\frac{i}{p\!\!\!/+k\!\!\!/_2-m_d}
i\gamma^{\rho}\gamma^{5}\frac{i}{p\!\!\!/-k\!\!\!/_1-m_s}(1-\gamma^5)\gamma^{\mu}\right],
\eea
the final integration result is
\bea
\label{bbb3}
&&\frac{(2\pi)^2n}{\sqrt{(q^2-k_1^2-k_2^2)^2-4 k_1^2 k_2^2}}\left[(m_d-m_u)\left(m_s(\frac{1}{2}-\frac{m_s^2-m_u^2}{2k_1^2})+m_u(\frac{1}{2}+\frac{m_s^2-m_u^2}{2k_1^2})\right)\right.\nonumber \\
&&\left.-(m_d+m_u)\left(-m_s(\frac{1}{2}-\frac{m_s^2-m_u^2}{2k_1^2})+m_u(\frac{1}{2}+\frac{m_s^2-m_u^2}{2k_1^2})\right)\right.\nonumber \\
&&\left.+(m_u-m_s)\left(m_s(\frac{1}{2}-\frac{m_s^2-m_u^2}{2k_2^2})-m_u(\frac{1}{2}+\frac{m_s^2-m_u^2}{2k_2^2})\right)\right.\nonumber \\
&&\left.-(m_u+m_s)\left(m_s(\frac{1}{2}-\frac{m_s^2-m_u^2}{2k_2^2})+m_u(\frac{1}{2}+\frac{m_s^2-m_u^2}{2k_2^2})\right)\right]\nonumber \\
&&=(m_d-m_u)J(m_u,-m_d,m_s)-(m_d+m_u)J(m_u,m_d,-m_s)\nonumber \\
&&+(m_u-m_s)J(-m_u,-m_d,m_s)-(m_u+m_s)J(m_u,-m_d,m_s).
\eea
 The masses of $K^0$ and $\bar{K}^0$ can be obtained from the equations (\ref{bbb1}), (\ref{bbb2}) and (\ref{bbb3}).

\section{Comparing with the experimental values}\label{ComparingWithExperimentalValues}

In a simplified picture mesons are considered as quark-anti-quark bound states, they have both radial and angular motion modes.
We study the masses of ground states which are corresponding to quantum number $n=1$. The constitute quark mass parameters used in the calculation are taken values $m_u=15.42\ {\rm{MeV}}$, $m_d=16.08\ {\rm{MeV}}$ and $m_s=121\ {\rm{MeV}}$.

From equation (\ref{qqu}), we can obtain the mass of $\pi^0$ by the relation $q^2=m_{\pi^0}^2$, the external momentums of photons satisfy the on-shell conditions $k_1^2=k_2^2=0$. The quark content of $\pi^0$ is $(u\bar{u}-d\bar{d})/\sqrt{2}$,
so the coefficients $\alpha$, $\beta$ and $\gamma$ in equation (\ref{qqu}) are
$\alpha=\frac{1}{\sqrt{2}}, \beta=-\frac{1}{\sqrt{2}},\gamma=0$.
Then we get
\bea
m_{\pi^0}=\sqrt{8\pi^2\frac{4m_u^2-m_d^2}{3}}\approx 135 \ {\rm{MeV}}.
\eea

\begin{figure}[H]
\center
\includegraphics[width=0.8\textwidth]{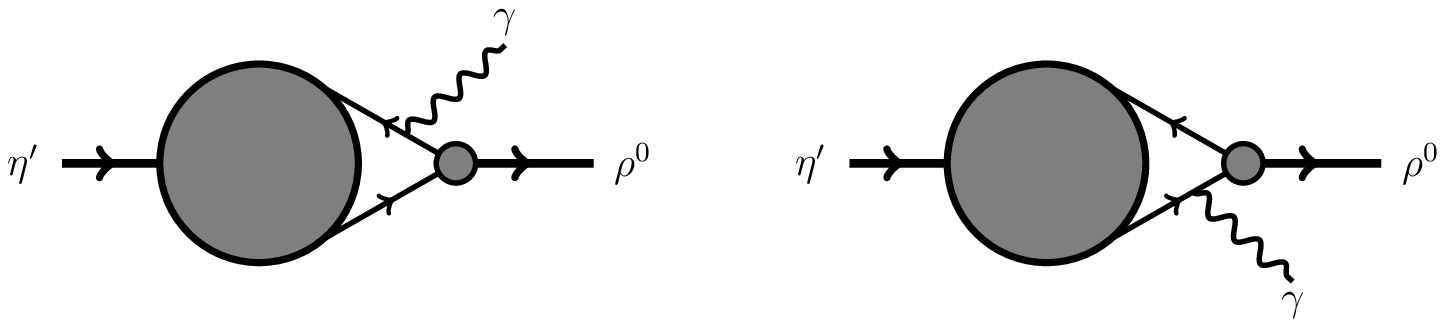}
\caption{\label{etaprimedecay} The decay channels of $\eta^{\prime}\to\rho^0\gamma$.  }
\end{figure}

The particles $\eta$ and $\eta'$ are mixing between the $\eta_0$ and $\eta_8$ states \cite{Holstein:2001bt}, with
$(u\bar{u}+d\bar{d}+s\bar{s})/\sqrt{3}$ for $\eta_0$ and $(u\bar{u}+d\bar{d}-2s\bar{s})/\sqrt{6}$ for $\eta_8$.
Without mixing, one can take $\eta$ as the $\eta_0$ and $\eta'$ as the $\eta_8$. The main anomalous decay mode of the $\eta'$  is $\eta'\rightarrow \rho^0 \gamma$ \cite{Tanabashi:2018oca}, as shown in Fig. \ref{etaprimedecay}.
From equation (\ref{qqu}), using the coefficients $\alpha, \beta, \gamma$ inferred from their flavor components, we obtain the mass of $\eta_0$ and $\eta_8$
\bea
&&m_{\eta_0}=\sqrt{8\pi^2\frac{4m_u^2+m_d^2+m_s^2}{6}}\approx 457 \ {\rm{MeV}},\\
&&m_{\eta_8}=\sqrt{-8\pi^2\frac{4m_u^2+m_d^2-2m_s^2}{3}+m_{\rho^0}^2}\approx 1157 \ {\rm{MeV}},
\eea
where the $m_{\rho^0}\approx 775 \ {\rm{MeV}}$ is the mass of vector meson $\rho^0$ \cite{Tanabashi:2018oca}.
The pseudoscalar mixing angle is $\theta\approx-20^{\circ}$ \cite{Holstein:2001bt}, then we obtain
\bea
&&m_{\eta}=\cos^2\theta m_{\eta_0}+\sin^2\theta m_{\eta_8}\approx 539 \ {\rm{MeV}},\\
&&m_{\eta'}=\sin^2\theta m_{\eta_0}+\cos^2\theta m_{\eta_8}\approx 1075 \ {\rm{MeV}}.
\eea

The masses of $\pi^{\pm}$  and $K^{\pm}$ can be obtained by solving the equations (\ref{aaa222}) and (\ref{aaa22222}). The result is
\bea
m_{\pi^{\pm}}\approx 140 \ {\rm{MeV}}, \quad
m_{K^{\pm}}\approx 507 \ {\rm{MeV}}.
\eea

The masses of $K^0$ and $\bar{K}^0$ can be calculated from two formulae, since there are two ways to manipulate the order of gamma matrices, explained in section \ref{MassesKandKBAR}. By the single projector prescription, from the equations (\ref{aaa31}), (\ref{aaa32}) and (\ref{aaa33}) we obtain
\bea
m_{K^0}=m_{\bar{K}^0}\approx 471 \ {\rm{MeV}}
\eea
By the normal Pauli-Villars prescription, from the equations (\ref{bbb1}), (\ref{bbb2}) and (\ref{bbb3}), we obtain
\bea
m_{K^0}=m_{\bar{K}^0}\approx 322 \ {\rm{MeV}}.
\eea
The result of single projector prescription agrees better with the experimental value.

We compare our results with the experimental values in Table \ref{tabi1}.
In general, the theoretical value of masses match better with experimental values for lighter mesons, if we consider the renormalization of quark masses which decrease with increasing energy scale, the match for $K^{\pm}$ and $\eta'$ can be improved.
It is interesting that, while the correct prescription for the $\gamma^5$ matrix in chiral theories is undetermined at the moment, here we need to adopt the single projector prescription for the $\gamma^5$ matrix to achieve a reasonably good theoretical masses for $K^0, \bar{K}^0$.
\begin{table}[htbp]
  \centering
  \begin{tabular}{c|c|c|c|c|c|c}
\hline

Pseudoscalar meson &  $\pi^0$ &$\pi^{\pm}$ & $K^{\pm}$ & $K^0, \bar{K}^0$ & $\eta$  &  $\eta'$   \\

\hline

 mass (experimental) & 135 MeV &140 MeV & 494 MeV & 498 MeV & 548 MeV  &  958 MeV   \\

\hline

 mass (theoretical) & 135 MeV &140 MeV & 507 MeV & 471 MeV & 539 MeV  &  1075 MeV   \\

\hline
  \end{tabular}
  \caption{\label{tabi1} Comparing with the experimental values which are taken from \cite{Tanabashi:2018oca}.}
  \end{table}

\section{Conclusions and Discussions}

In this paper, we have studied the masses of light pseudoscalar mesons by a new method based on the previous study \cite{Liu:2021pmt,Liu:2022zdo}. The integral formula (\ref{fz}) for the multi-valued complex function is applied to perform the integrals arise in the amplitudes of triangle diagrams associated with the axial vector current, the integration results allows to cancel the ABJ anomaly term, therefore eliminate the anomalous problem. This new anomaly free condition imposes a relation between the mass of meson and the mass its constitute quarks by which the mass of pseudoscalar particle can be determined. Our results agree well with experiments very well. To obtain more accurate results, we need to consider the  renormalization group running of quark mass through the renormalization group equations, this would be considered in future work.


\section*{Acknowledgements}
This work is supported by Chinese Universities Scientific Fund Grant No. 2452018158, by a grant from CWNU (No. 18Q068).

\end{document}